\documentclass[11pt]{article}

\usepackage[final]{acl}

\usepackage{times}
\usepackage{latexsym}

\usepackage[T1]{fontenc}

\usepackage[utf8]{inputenc}

\usepackage{microtype}

\usepackage{inconsolata}

\usepackage{graphicx}

\usepackage{booktabs}   
\usepackage{amssymb}
\usepackage{amsmath}
\usepackage{multirow} 

\title{CreditXAI: A Multi-Agent System for Explainable Corporate Credit Rating}

\author{
  Yumeng Shi\textsuperscript{1},\quad 
  Zhongliang Yang*\textsuperscript{1},\quad 
  Yisi Wang\textsuperscript{2},\quad 
  Linna Zhou\textsuperscript{1} \\
  \textsuperscript{1}School of Cyberspace Security, Beijing University of Posts and Telecommunications, Beijing, China \\
  \textsuperscript{2}Guotai Junan Securities, Shanghai, China \\
  \texttt{yangzl@bupt.edu.cn}
}

\begin{document}
\maketitle
\begin{abstract}
In the domain of corporate credit rating, traditional deep learning methods have improved predictive accuracy but still suffer from the inherent 'black-box' problem and limited interpretability. While incorporating non-financial information enriches the data and provides partial interpretability, the models still lack hierarchical reasoning mechanisms, limiting their comprehensive analytical capabilities. To address these challenges, we propose \textbf{CreditXAI}, a Multi-Agent System (MAS) framework that simulates the collaborative decision-making process of professional credit analysts. The framework focuses on business, financial, and governance risk dimensions to generate consistent and interpretable credit assessments. Experimental results demonstrate that multi-agent collaboration improves predictive accuracy by more than 7\% over the best single-agent baseline, confirming its significant synergistic advantage in corporate credit risk evaluation. This study provides a new technical pathway to build intelligent and interpretable credit rating models.
\end{abstract}

\section{Introduction}

Corporate credit rating is a crucial component of the modern financial system, essential for market stability and efficient capital allocation \citep{A1-hilscher2017credit}. Methodologies have evolved from traditional statistical and machine learning models \citep{A2-friedman1991multivariate,A3-altman1977zetatm,A4-huang2004credit,A5-yeh2012hybrid,A6-abellan2017comparative} to a paradigm shift driven by deep learning architectures such as CNNs, GNNs, and Transformers \citep{A7-feng2020every,A8-chen2020novel,A9-feng2022every,A20-tavakoli2025multi}. 
Recognizing that purely financial metrics provide an incomplete picture, researchers have increasingly turned to non-financial data, such as corporate 10-K reports, to enrich the analysis. This integration also offers a potential pathway to mitigate the 'black-box' interpretability issues of advanced models \citep{A11-balakrishnan2010predictive}. Although these reports provide authoritative, rich information, their high dimensionality, sparsity, and semantic heterogeneity present formidable challenges for fine-grained feature extraction and effective multimodal fusion.

\begin{table}[t]
\begin{minipage}{0.95\columnwidth}
\raggedleft
\caption{Comparison of Representative Methods for Corporate Credit Rating}
\label{tab:comparison}
\renewcommand{\arraystretch}{1.0}
\resizebox{\textwidth}{!}{
\begin{tabular}{lccc}
\toprule
\textbf{Method} & \textbf{Model} & \textbf{Non-fin} & \textbf{XAI} \\
\midrule
\citep{A17-golbayani2020application} & CNN & $\times$ & $\times$ \\
\citep{A7-feng2020every} & CNN & $\times$ & $\times$ \\
\citep{A18-feng2021adversarial} & CNN + Adv. & $\times$ & $\times$ \\
\citep{A19-feng2022contrastive} & CNN + Contr. & $\times$ & $\times$ \\
\citep{A9-feng2022every} & GNN & $\triangle$ & $\times$ \\
\citep{A20-tavakoli2025multi} & Transf. & $\times$ & $\triangle$ \\
\citep{A21-shi2024sparsegraphsage} & GNN & $\times$ & $\times$ \\
\citep{A23-choi2020predicting} & ML (BOW/Word2Vec/Doc2Vec) & $\checkmark$ & $\times$ \\
\citep{A24-zhang2023investment} & CatBoost / LGBM & $\checkmark$ & $\triangle$ \\
\citep{A25-chen2024social} & KNN & $\checkmark$ & $\triangle$ \\
\citep{A12-shi2025creditarf} & Transf. & $\checkmark$ & $\times$ \\
\citep{A13-tan2025explainable} & Transf. & $\checkmark$ & $\checkmark$ \\
\textbf{Ours (CreditXAI)} & \textbf{Agentic System (LLM-based)} & \textbf{$\checkmark$} & \textbf{$\checkmark$} \\
\bottomrule
\end{tabular}}
\vspace{1mm}
\footnotesize{\textit{Note:} $\checkmark$ = full, $\times$ = none, $\triangle$ = partial use.}
\end{minipage}
\end{table}

Current monolithic Large Language Model (LLM) approaches to credit rating lack nuanced, hierarchical reasoning and suffer from poor interpretability \citep{A12-shi2025creditarf}. Similarly, while the XAI frameworks \citep{A13-tan2025explainable} offer valuable risk visualization, they still do not address the core challenge of emulating the structured reasoning process of expert analysis. Multi-Agent Systems (MAS) offer a compelling alternative, simulating expert collaboration to potentially achieve both high accuracy and transparency \citep{A14-yu2025table,A15-yu2024fincon,A16-jajoo2025masca}. However, the application of MAS to corporate credit rating remains largely unexplored. To bridge this research gap, we introduce \textbf{CreditXAI}, a novel framework leveraging multi-agent collaboration for accurate and interpretable credit risk analysis.

\begin{figure}[ht]
    \centering
    \includegraphics[width=\columnwidth]{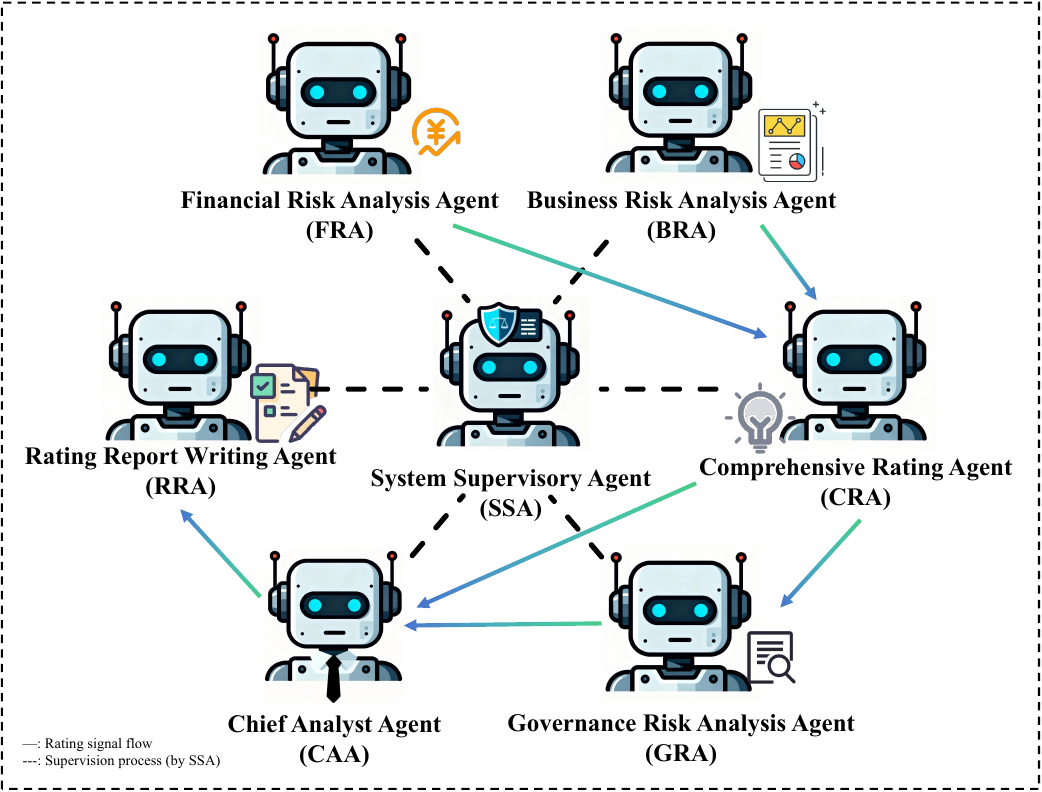}
    \caption{Overview of the \textbf{CreditXAI} framework. This multi-agent system is organized into four layers of specialized agents: the Analytical Layer (BRA, FRA, GRA, CRA), the Decision Layer (CAA, RRA), and the Supervisory Layer (SSA), all collaborating to generate explainable corporate credit ratings.}
    \label{fig:CreditXAI-agents}
\end{figure}

As depicted in Figure~\ref{fig:CreditXAI-agents}, the \textbf{CreditXAI} framework operates by emulating a professional team of credit analysts, where each agent embodies a specialized analytical role. These seven agents constitute a hierarchical and collaborative architecture designed to replicate the complex reasoning processes of human experts. Through coordinated analysis and multimodal information fusion, our framework produces robust, explainable, and traceable corporate credit ratings with high predictive accuracy. The primary contributions of this work are summarized as follows:

\begin{enumerate}
    \item \textbf{Designing a Multi-Agent Credit Rating Architecture with Specialized Roles}:
    We propose a modular and scalable multi-agent credit rating system, comprising agents specialized in business, financial, and governance risk analysis, as well as a composite rater and a chief analyst. This architecture supports inter-agent collaboration and dynamic information integration, enhancing the stability and reliability of credit ratings.
    \item \textbf{A Fine-Grained Mechanism for Unstructured Text Analysis}:
    Our framework enables agents to independently dissect individual items within 10-K annual reports, extracting nuanced semantic insights from unstructured and semistructured textual information. This mechanism allows for adaptive weighting of different risk dimensions based on the significance of extracted signals.
    \item \textbf{A History-Aware Framework for Intelligent Reasoning}:
    We develop a reasoning framework that uses historical time series data for decision making. By integrating semantic similarity calculations with a weighted reference mechanism, the system dynamically incorporates past ratings as benchmarks, enabling more robust and context-aware assessments.
\end{enumerate}

\begin{figure*}[t]
    \centering
    \includegraphics[width=\textwidth]{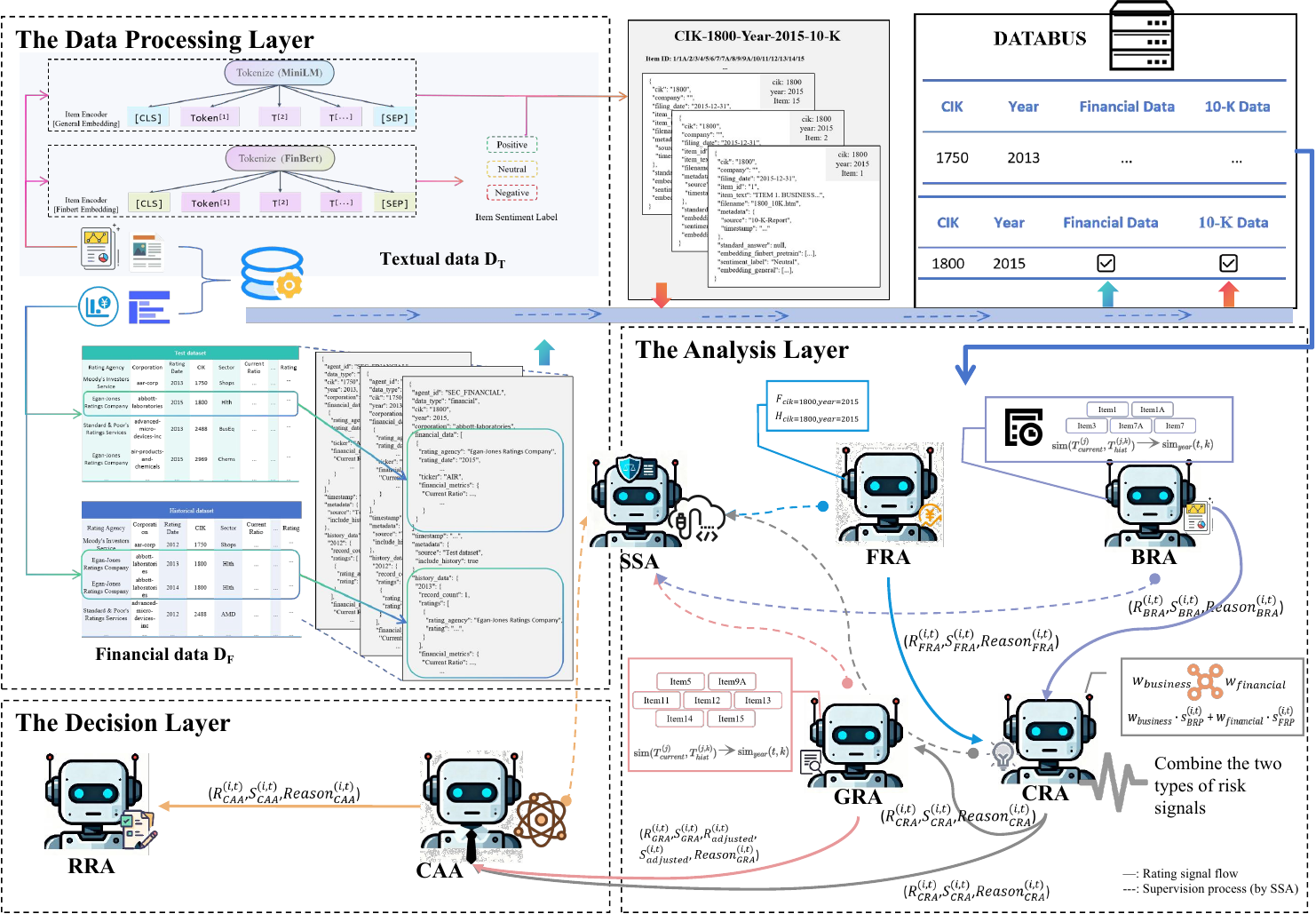}
    \caption{Overview of the \textbf{CreditXAI} framework.}
    \label{fig:framework}
\end{figure*}

\section{Related Work}

Research in corporate credit rating has evolved substantially over the past decade, progressing from traditional statistical models to sophisticated deep learning architectures and, more recently, toward multimodal and explainable AI. The early methodologies were dominated by statistical and machine learning techniques such as Logit regression, discriminant analysis, and Support Vector Machines \citep{A2-friedman1991multivariate, A3-altman1977zetatm}. The advent of deep learning instigated a paradigm shift, introducing architectures like Convolutional Neural Networks (CNNs), Recurrent Neural Networks (RNNs), Graph Neural Networks (GNNs), and Transformers to the field. Golbayani et al. (2020) were among the first to apply CNNs to this task, followed by a series of innovations that used two-dimensional financial visualizations, adversarial learning, and contrastive pre-training to enhance feature representation and robustness \citep{A7-feng2020every, A18-feng2021adversarial, A19-feng2022contrastive}. To model company-to-company dependencies through equity ties, supply chain connections, and industry relationships, researchers employed GNNs \citep{A9-feng2022every,A21-shi2024sparsegraphsage}, while others utilized multi-task Transformers to facilitate knowledge sharing across related tasks \citep{A20-tavakoli2025multi}. Despite their success with structured financial data, these models often exhibit a limited capacity for integrating complex, non-financial information.

To address this limitation, recent research has increasingly focused on incorporating non-financial textual data to enrich credit risk models. Studies have used information from diverse sources, including social media sentiment \citep{A22-yuan2018mining}, business descriptions from annual reports \citep{A24-zhang2023investment}, and sentiment-driven textual analysis \citep{A23-choi2020predicting,A25-chen2024social}. In parallel, the critical demand for transparency has spurred the development of Explainable AI (XAI) techniques. Frameworks such as TinyXRA have sought to embed interpretability in model design by using lightweight Transformers and attention visualization \citep{A13-tan2025explainable}. However, these approaches, being monolithic, inherently process information through a unified reasoning path. This often leads to an oversight of granular document-level structures and an inability to emulate the collaborative, multi-faceted reasoning characteristic of human financial analysis, thereby limiting their reliability in high-stakes applications.

Multi-Agent Systems (MAS) present a promising paradigm to overcome the shortcomings of monolithic architectures. As distributed intelligent systems, MAS employ a consortium of autonomous agents that collaborate to solve complex tasks, mimicking the division of labor within an expert team. They have demonstrated notable potential in adjacent financial domains, including stock market prediction \citep{A15-yu2024fincon} and personal credit risk analysis \citep{A16-jajoo2025masca}. 

However, systematic exploration of Multi-Agent Systems (MAS) for corporate credit rating applications remains limited in the literature, particularly for tasks involving synergistic reasoning over both structured and unstructured data. As the comparative analysis in Table~\ref{tab:comparison} illustrates, 'Model,' 'Non-fin,' and 'XAI,' respectively, refer to the model type, use of non-financial information, and explainability level. Existing research in corporate credit rating has focused predominantly on monolithic architectures, with interpretability often relying on post hoc techniques, and the fusion of multimodal information remains to be systematically addressed.

Against this backdrop, applying the MAS paradigm to corporate credit rating for synergistic reasoning remains a promising yet underexplored direction. To bridge this gap, this paper proposes \textbf{CreditXAI}, a framework that integrates the collaborative intelligence of MAS with the semantic capabilities of Large Language Models (LLMs). By assigning specialized agent roles and enabling semantic collaboration with dynamic weight fusion, \textbf{CreditXAI} provides process-level explainability and traceability in credit rating decisions. The framework offers clear advantages in predictive performance, transparency, and scalability, establishing a new paradigm for intelligent and interpretable credit ratings.

\section{Proposed Method}
\label{sec:proposed_method}

We introduce \textbf{CreditXAI}, a novel Multi-Agent System (MAS) framework for explainable corporate credit ratings. By emulating a human analysis team through a modular, hierarchical architecture, our framework delivers a multifaceted risk analysis with an interpretable and traceable decision-making process (see Figure~\ref{fig:framework}).

\subsection{System Architecture}

The \textbf{CreditXAI} architecture is organized into four distinct layers: a Data Processing Layer, an Analysis Layer, a Decision Layer, and a Supervision Layer. For a given company $i$ in year $t$, the system takes as input a raw dataset $D^{(i,t)} = \{D_F^{(i,t)}, D_T^{(i,t)}\}$, where $D_F^{(i,t)}$ represents structured financial data and $D_T^{(i,t)}$ denotes unstructured textual data from sources such as 10-K reports.

The \textbf{Data Processing Layer} is responsible for transforming this raw, heterogeneous data into a unified and structured feature representation, denoted as $\mathcal{D}_{\text{processed}}^{(i,t)}$. This processed data then serves as the input for the \textbf{Analysis Layer}, which comprises four specialized agents. These agents are tasked with generating distinct risk signals, specifically the Business Risk Profile ($R_{\text{BRA}}, S_{\text{BRA}}$), the Financial Risk Profile ($R_{\text{FRA}}, S_{\text{FRA}}$), a Composite Rating ($R_{\text{CRA}}, S_{\text{CRA}}$), and the Governance Risk Profile ($R_{\text{GRA}}, S_{\text{GRA}}$). Each agent outputs both a categorical rating \(R \in \{\text{AAA}, \dots, \text{C}\}\) and a continuous risk score \(S \in [0,1]\), which can be converted to one another via a unified mapping. Agent-specific computations are defined as follows:
\begin{equation}
\label{eq:analysis_layer_funcs}
\begin{aligned}
R_{\text{BRA}}, S_{\text{BRA}} &= f_{\text{BRA}}(\mathcal{D}_{\text{T}}), \\
R_{\text{FRA}}, S_{\text{FRA}} &= f_{\text{FRA}}(\mathcal{D}_{\text{F}}), \\
R_{\text{CRA}}, S_{\text{CRA}} &= f_{\text{CRA}}(S_{\text{BRA}}, S_{\text{FRA}}), \\
R_{\text{GRA}}, S_{\text{GRA}} &= f_{\text{GRA}}(\mathcal{D}_{\text{T}}).
\end{aligned}
\end{equation}

Finally, the \textbf{decision layer} dynamically aggregates these individual risk profiles. Its Chief Analyst Agent (CAA) agent produces the final synthesized risk score and rating:
\begin{equation}
\label{eq:decision_layer_funcs}
R_{\text{CAA}}, S_{\text{CAA}} = f_{\text{CAA}}(R_{\text{CRA}}, S_{\text{CRA}}, R_{\text{GRA}}, S_{\text{GRA}}).
\end{equation}
The final outputs of the framework are \(S_{\text{final}} = S_{\text{CAA}}\) and \(R_{\text{final}} = R_{\text{CAA}}\).

\subsection{Data Processing Layer}

\paragraph{Financial Data Processing.} This layer preprocesses and standardizes structured financial data from SEC filings. It constructs a time series dataset for each company by defining a historical window of length $K$. This historical context, $H_{i,t}$, encapsulates past financial statements $F$ and their corresponding ratings $R$ over the period $\{t-K, \dots, t-1\}$:
{\small
\begin{equation}
\label{eq:historical_data}
H_{i,t} = \{(F_{i,t-K}, R_{i,t-K}), \dots, (F_{i,t-1}, R_{i,t-1})\}.
\end{equation}
}

\paragraph{10-K Report Data Processing.}
This layer also parses 10-K reports to extract key sections (e.g., \emph{Item 1A: Risk Factors}, \emph{Item 7: MD\&A}). For each item $j$, it constructs a hybrid semantic representation by generating:
(1) financial domain-specific embeddings $\mathbf{v}_{\text{finbert}}$,
(2) general semantic embeddings $\mathbf{v}_{\text{general}}$,
and (3) a sentiment score $item_{\text{sentiment}}$.
The resulting feature set for the $j$-th item of the company $i$ in year $t$ is defined as:
\begin{equation}
\label{eq:item_representation}
S_{i,t}^{(j)} =
\left\{
\mathbf{v}_{\text{finbert}}^{(j)},\;
\mathbf{v}_{\text{general}}^{(j)},\;
item_{\text{sentiment}}^{(j)}
\right\}.
\end{equation}

\subsection{Analysis Layer}

The Analysis Layer is the core of the risk assessment process, comprising four distinct analytical agents.

\paragraph{Business Risk Analysis Agent (BRA).} The BRA evaluates the company's business risks by analyzing the semantic content of core 10-K items. It incorporates a historically informed approach by leveraging historical data. Given the embedding vector for a current item $j$, $\mathbf{v}_{i,t}^{(j)}$, and its historical counterpart from year $t-k$, $\mathbf{v}_{i,t-k}^{(j)}$, the agent computes the cosine similarity to measure semantic drift:
\begin{equation}
\label{eq:semantic_similarity}
\text{sim}_{j,k} =
\frac{\mathbf{v}_{i,t}^{(j)} \cdot \mathbf{v}_{i,t-k}^{(j)}}{\|\mathbf{v}_{i,t}^{(j)}\| \cdot \|\mathbf{v}_{i,t-k}^{(j)}\|}.
\end{equation}

Historical ratings $R_{i,t-k}$ are then weighted based on the aggregated semantic similarity $\text{sim}_k$ (derived from Eq.~\ref{eq:semantic_similarity}) between the reports from year $t$ and year $t-k$, normalized via a softmax function:
\begin{equation}
\label{eq:historical_weights}
w_k = \frac{\exp(\alpha \cdot \text{sim}_{k})}{\sum_{l=1}^{K} \exp(\alpha \cdot \text{sim}_{l})}.
\end{equation}

Rather than direct numerical aggregation, these weights guide the agent to emphasize historically similar years when interpreting the current disclosure. The agent then produces an updated business rating $R_{\text{BRA}}$, which is mapped to a continuous risk score $S_{\text{BRA}}$ for downstream fusion.

\paragraph{Financial Risk Analysis Agent (FRA).} The FRA is responsible for the quantitative assessment of a company's financial risk. Its evaluation integrates both cross-sectional industry benchmarking and longitudinal time series analysis.

First, it establishes industry-specific baseline values $\text{baseline}_{s,j}$ for each financial indicator $j$ 
in the sector $s$ by calculating the median values from historical industry data:
\begin{equation}
\label{eq:baseline_calculation}
\text{baseline}_{s,j} = \operatorname{median}\!\bigl(f_{k,t}^{(j)}\bigr),
\end{equation}
where the median is computed over all companys $k \in \text{sector}_s$ 
and across all historical periods $t \in \text{historical\_periods}$ included in the dataset.

Second, the deviation of each financial indicator $f_{i,t}^{(j)}$ from its industry benchmark is calculated:
\begin{equation}
\label{eq:deviation_fra}
\text{dev}_{i,j}^{(t)} = \frac{f_{i,t}^{(j)} - \text{baseline}_{s,j}}{\text{baseline}_{s,j}}.
\end{equation}

Third, the agent evaluates each financial indicator relative to the company's historical rating and decides, based on year-over-year changes and calibrated thresholds, whether to adjust it. It outputs the updated categorical rating \(R_{\text{FRA}}\) with a rationale, which is then mapped to a continuous financial risk score \(S_{\text{FRA}}\). Both \(R_{\text{FRA}}\) and \(S_{\text{FRA}}\) are stored in the financial risk signal.

\paragraph{Comprehensive Rating Agent (CRA).} The CRA combines business and financial risk scores into a unified composite rating. It employs a dynamic weighting mechanism where the weights $w_{\text{BRA}}, w_{\text{FRA}}$ are adjusted based on the absolute difference $\Delta S = |S_{\text{BRA}} - S_{\text{FRA}}|$ relative to a threshold $\delta$:
\begin{equation}
\label{eq:CRA_rating}
S_{\text{CRA}} = w_{\text{BRA}} \cdot S_{\text{BRA}} + w_{\text{FRA}} \cdot S_{\text{FRA}},
\end{equation}
where $w_{\text{BRA}}$ is assigned a higher value ($w_{\text{high}}$) if the scores diverge significantly ($\Delta S > \delta$), and $w_{\text{FRA}} = 1 - w_{\text{BRA}}$.

\paragraph{Governance Risk Analysis Agent (GRA).} The GRA analyzes governance-related 10-K disclosures under a historical learning framework similar to the BRA, where semantically weighted historical information is fused with current textual evidence to produce the initial governance rating \(R_{\text{GRA}}\).

Distinct from the BRA, the GRA further determines a recommended adjustment based on its governance assessment. After receiving the composite rating \(R_{\text{CRA}}\) from the CRA, the GRA evaluates whether a governance-based adjustment is warranted and updates the initial governance rating accordingly, producing the adjusted governance rating \(R_{\text{adjusted}}\). Both \(R_{\text{GRA}}\) and \(R_{\text{adjusted}}\) are then provided as governance risk signals for subsequent fusion and decision-making.

\subsection{Decision Layer}

\paragraph{Chief Analyst Agent (CAA).} 
The CAA synthesizes quantitative risk scores ($S_{\text{FRA}}, S_{\text{BRA}}, S_{\text{GRA}}, S_{\text{CRA}}$), categorical ratings ($R_{\text{FRA}}, R_{\text{BRA}}, R_{\text{GRA}}, R_{\text{CRA}}$), and supporting rationale to produce a final aggregated risk score $S_{\text{CAA}}$ and an overall rating $R_{\text{CAA}}$. By evaluating the consistency and reliability of these signals, the CAA adjusts their influence accordingly, ensuring a comprehensive, consensus-aware, and interpretable credit assessment.

\paragraph{Rating Report Writing Agent (RRA).} 
The RRA generates a structured, human-readable report based on the CAA's outputs (\(S_{\text{CAA}}\), \(R_{\text{CAA}}\)), integrating quantitative scores and qualitative reasoning from business, financial, governance, and composite analyses into a concise, interpretable, and actionable summary.

\subsection{Supervisory Layer}
The Supervisory Layer oversees the operations of all analytical agents within \textbf{CreditXAI}. Its core component is the \textbf{System Supervisory Agent (SSA)}, which leverages the LangSmith architecture to monitor agent inputs, outputs, and information flows. 

\section{Experimental Results}

\paragraph{Experimental Setup.}
We use 5,403 company-year samples from U.S. companies\footnote{\url{https://www.kaggle.com/datasets/kirtandelwadia/corporate-credit-rating-with-financial-ratios}}, filtered for completeness and split into a historical reference set for agent learning and a test set for evaluation. \textbf{CreditXAI} leverages multiple financial dimensions and textual embeddings from 14 key 10-K items (FinBERT\footnote{\url{https://huggingface.co/yiyanghkust/finbert-pretrain}}, MiniLM\footnote{\url{https://huggingface.co/sentence-transformers/all-MiniLM-L6-v2}}, FinBERT-Tone\footnote{\url{https://huggingface.co/yiyanghkust/finbert-tone}}) capturing content and sentiment, with ground-truth ratings from major agencies. We compare against \textbf{CreditARF} baselines \citep{A12-shi2025creditarf} using accuracy, recall, and F1 score, evaluating historical learning via a \textit{History Group} (progressive learning) versus a \textit{No-History Group} (current-year only), complemented by ablation studies. The framework is built on LangChain, with LangGraph orchestrating agents, LangSmith ensuring traceability, and each agent powered by \textbf{Gemini 2.0 Flash}.

\subsection{Experimental Results Analysis}

\subsubsection{Baseline Model Performance}

\begin{table}[ht]
\centering
\small
\caption{Performance comparison of baseline models and \textbf{CreditXAI} agents on different input types. ``Imp.'' indicates relative improvement or decline compared with the best baseline.}
\label{tab:baseline-performance}
\setlength{\tabcolsep}{2pt} 
\resizebox{\columnwidth}{!}{%
\begin{tabular}{c c c c c c c} 
\toprule
Data & Model & Work & ACC & F1-Score & Recall & Imp. \\
\midrule
\multirow{3}{*}{10-K} & MLP & \citep{A12-shi2025creditarf} & 0.235 & 0.128 & 0.235 & Baseline\\
                       & Agent & \textbf{Ours(BRA)}& \textbf{0.541} & \textbf{0.598}& \textbf{0.541}& \textbf{$\uparrow$ 30.6\%}\\
                       & Agent & \textbf{Ours(GRA)}& \textbf{0.563} & \textbf{0.564}& \textbf{0.563}& \textbf{$\uparrow$ 32.8\%} \\
\midrule
\multirow{4}{*}{Fin} & CNN & \citep{A7-feng2020every} & 0.522 & 0.507 & 0.522 & - \\
                      & GNN & \citep{A9-feng2022every} & 0.522 & 0.427 & 0.522 & Baseline\\
                      & LSTM & \citep{A20-tavakoli2025multi} & 0.349 & 0.242 & 0.349 & -\\
                      & Agent & \textbf{Ours(FRA)}& \textbf{0.647} & \textbf{0.657}& \textbf{0.647}& \textbf{$\uparrow$ 12.5\%} \\
\midrule
\multirow{5}{*}{Fin+10-K} & CNN & \citep{A12-shi2025creditarf} & 0.625 & 0.619 & 0.625 & -\\
                           & GNN & \citep{A12-shi2025creditarf} & 0.656 & 0.653 & 0.656 & Baseline\\
                           & LSTM & \citep{A12-shi2025creditarf} & 0.651 & 0.649 & 0.651 & -\\
                           & MAS & \textbf{Ours(CRA)}& \textbf{0.713} & \textbf{0.712}& \textbf{0.713} & \textbf{$\uparrow$ 5.7\%} \\
                           & \textbf{MAS} & \textbf{Ours(CAA)}& \textbf{0.726} & \textbf{0.727}& \textbf{0.726} & \textbf{$\uparrow$ 7\%} \\
\bottomrule
\end{tabular}%
}
\end{table}

Table~\ref{tab:baseline-performance} shows that \textbf{CreditXAI} consistently outperforms the best baseline models across all data types (10-K text, financial data, and their fusion) under identical input conditions. In the textual domain, the \textbf{BRA} and \textbf{GRA} achieve improvements of 30.6\% and 32.8\%, respectively, over the corresponding baselines \citep{A12-shi2025creditarf}. For financial data, the \textbf{FRA} surpasses several deep learning baselines \citep{A7-feng2020every, A9-feng2022every, A20-tavakoli2025multi} with a gain of 12.5\%. When text and financial data are combined, the \textbf{CRA} and \textbf{CAA} improve by 5.7\% and 7\%, respectively. These results confirm that \textbf{CreditXAI} enhances generalization and robustness through multi-agent collaboration and historical learning.

\subsubsection{Validating the Historical Learning Strategy}

\begin{table}[ht]
\centering
\footnotesize 
\caption{Performance improvement of agents with historical learning. ``Imp.'' indicates the improvement of accuracy compared with the baseline.}
\label{tab:historical-learning}
\setlength{\tabcolsep}{3pt} 
\begin{tabular}{c c c c}
\toprule
Agent & ACC (Baseline) & ACC (Historical) & Imp. \\
\midrule
BRA & 0.013 & \textbf{0.541} & \textbf{$\uparrow$ 52.8\%} \\
FRA & 0.284 & \textbf{0.647} & \textbf{$\uparrow$ 36.3\%} \\
CRA & 0.315 & \textbf{0.713} & \textbf{$\uparrow$ 39.8\%} \\
GRA & 0.320 & \textbf{0.563} & \textbf{$\uparrow$ 24.3\%} \\
\bottomrule
\end{tabular}
\end{table}

An ablation study (Table~\ref{tab:historical-learning}) confirms that historical learning is critical for robust analysis. Access to past data enabled the \textbf{BRA} agent's accuracy to soar from 0.013 to 0.541. Similarly, the \textbf{FRA}, \textbf{CRA}, and \textbf{GRA} achieved substantial accuracy gains of 36.3\%, 39.8\%, and 24.3\%, respectively. From an agentic engineering perspective, these results validate our stateful agents, which use historical data to progressively make more informed predictions.

\begin{table*}[ht]
\centering
\scriptsize
\caption{Performance metrics of standalone agents across credit rating categories. AAA–C represent different credit ratings from highest to lowest. “ACC” denotes per-class Precision, while “Overall ACC” represents overall Accuracy across all categories. Metrics include Precision (ACC), Recall, and F1-Score.}
\label{tab:single-agent}
\setlength{\tabcolsep}{2pt}
\renewcommand{\arraystretch}{1.0}
\resizebox{1.0\textwidth}{!}{
\begin{tabular}{c c c c c c c c c c c}
\toprule
Agent & Metric & AAA & AA & A & BBB & BB & B & CCC & C & Overall \\
\midrule
\multirow{3}{*}{Business Risk Agent (BRA)} 
& ACC       & 1.000 & 0.800 & 0.726 & 0.787 & 0.672 & 0.565 & 0.219 & 0.016 & 0.541 \\
& Recall    & 0.500 & 0.364 & 0.570 & 0.573 & 0.500 & 0.557 & 0.636 & 0.500 & 0.541 \\
& F1-Score  & 0.667 & 0.500 & 0.638 & 0.663 & 0.573 & 0.561 & 0.326 & 0.032 & 0.598 \\
\midrule
\multirow{3}{*}{Financial Risk Agent (FRA)}
& ACC       & 0.800 & 0.562 & 0.747 & 0.757 & 0.656 & 0.594 & 0.269 & 0.333 & 0.647 \\
& Recall    & 0.400 & 0.409 & 0.709 & 0.736 & 0.656 & 0.543 & 0.636 & 0.500 & 0.647 \\
& F1-Score  & 0.533 & 0.474 & 0.727 & 0.747 & 0.656 & 0.567 & 0.378 & 0.400 & 0.657 \\
\midrule
\multirow{3}{*}{Governance Risk Agent (GRA)}
& ACC       & 1.000 & 0.769 & 0.658 & 0.539 & 0.554 & 0.750 & 0.250 & 1.000 & 0.563 \\
& Recall    & 0.400 & 0.455 & 0.316 & 0.627 & 0.800 & 0.514 & 0.455 & 0.500 & 0.563 \\
& F1-Score  & 0.571 & 0.571 & 0.427 & 0.580 & 0.655 & 0.610 & 0.323 & 0.667 & 0.564 \\
\bottomrule
\end{tabular}%
}
\end{table*}

\subsubsection{Validation of Agent Specialization}
Table~\ref{tab:single-agent} summarizes the performance of standalone agents in credit rating categories. Among BRA, FRA, and GRA, the \textbf{FRA} achieved the highest accuracy (0.647) on structured data, the \textbf{BRA} (0.541) excels at top companies from textual information, and the \textbf{GRA} (0.563) effectively flags extreme risks. These complementary strengths highlight that no single agent suffices, justifying the collaborative fusion in \textbf{CreditXAI}.

\subsubsection{Evaluation of Multi-Agent Collaboration}

The benefits of multi-agent collaboration are evident in Table~\ref{tab:multi-agent-fusion}, which compares overall accuracy and improvements from agent fusion strategies. The \textbf{CRA}, which combines the outputs of the \textbf{BRA} and \textbf{FRA}, achieves an accuracy of 0.713, a \textbf{6.6\%} improvement over the best single agent (FRA at 0.647), demonstrating the value of integrating multimodal data. Building on this, the \textbf{GRA} incorporates governance-related insights to refine the assessment, paving the way for the next hierarchical level. At this level, the \textbf{CAA} fuses all specialized outputs, including those from \textbf{GRA}, further increasing accuracy to 0.726, a cumulative \textbf{7.9\%} gain. These results illustrate how hierarchical collaboration of the \textbf{CreditXAI} progressively enhances predictive performance, robustness, and interpretability in corporate credit rating.

\begin{table}[ht]
\centering
\small
\caption{Performance of single-agent and multi-agent fusion strategies. ``Imp.'' indicates relative improvement compared with the best single-agent baseline. In the Agent column, ``+'' denotes a combination of multiple agents.}
\label{tab:multi-agent-fusion}
\resizebox{\columnwidth}{!}{%
\begin{tabular}{c c c c c c}
\toprule
Agent & ACC & Recall & F1-Score & Imp. \\
\midrule
BRA & 0.541 & 0.541 & 0.598 & - \\
FRA & 0.647 & 0.647 & 0.657 & Baseline \\
GRA & 0.563 & 0.563 & 0.564 & - \\
CRA (BRA+FRA) & 0.713 & 0.713 & 0.712 & \textbf{$\uparrow$ 6.6\%} \\
CAA (BRA+FRA+GRA) & 0.726 & 0.726 & 0.727 & \textbf{$\uparrow$ 7.9\%} \\
\bottomrule
\end{tabular}%
}
\end{table}

\section{Conclusion}

This paper introduced \textbf{CreditXAI}, a multi-agent framework for explainable corporate credit rating, leveraging LLM-based intelligent agents. By combining specialized agent roles, multimodal data fusion, and historical learning, the framework integrates business, financial, and governance analyses to produce interpretable ratings. Experimental results suggest that while individual agents provide useful insights, hierarchical fusion through the \textbf{CRA} and \textbf{CAA} can improve overall performance. These findings indicate that multi-agent collaboration and structured aggregation may enhance both the robustness and transparency of corporate credit assessments, providing a practical approach for deploying LLM-based agents in financial decision-making tasks.

\section{Limitations}
Although \textbf{CreditXAI} demonstrates promising performance, there remain opportunities for further refinement. Its modular architecture can accommodate additional data types, such as ESG reports, corporate sentiment, and news, potentially enhancing the breadth of analysis. Incorporating larger and more diverse datasets may also improve generalization across different corporate profiles. Future work on deploying \textbf{CreditXAI} in practical financial settings will help assess its robustness and inform guidelines for real-world implementation.


\bibliography{custom}




\end{document}